\documentclass[manuscript,screen,nonacm]{acmart}

\usepackage{enumitem}
\usepackage{svg}
\usepackage{subfigure}

\setlist[itemize]{leftmargin=15pt}
\AtBeginDocument{%
  }

\begin{document}

\title[The AI Genie Phenomenon and Three Types of AI Chatbot Addictions]{The AI Genie Phenomenon and Three Types of AI Chatbot Addiction: Escapist Roleplays, Pseudosocial Companions, and Epistemic Rabbit Holes}

\author{M. Karen Shen}
\email{shen021@ece.ubc.ca}
\affiliation{%
  \institution{University of British Columbia}
  \city{Vancouver}
  \country{Canada}
}

\author{Jessica Huang}
\email{jessh11@alumni.ubc.ca}
\affiliation{%
  \institution{Georgia Institute of Technology}
  \city{Atlanta}
  \country{USA}
}

\author{Olivia Liang}
\email{olilia26@student.ubc.ca}
\affiliation{%
  \institution{University of British Columbia}
  \city{Vancouver}
  \country{Canada}
}

\author{Ig-Jae Kim}
\email{drjay@kist.re.kr}
\affiliation{%
  \institution{Korea Institute of Science and Technology}
  \city{Seoul}
  \country{Korea}
}

\author{Dongwook Yoon}
\email{yoon@cs.ubc.ca}
\affiliation{%
  \institution{University of British Columbia}
  \city{Vancouver}
  \country{Canada}
}

\label{abs}
\begin{abstract}
Recent reports on generative AI chatbot use raise concerns about its addictive potential. An in-depth understanding is imperative to minimize risks, yet AI chatbot addiction remains poorly understood. This study examines how to characterize AI chatbot addiction---why users become addicted, the symptoms commonly reported, and the distinct types it comprises. We conducted a thematic analysis of Reddit entries (n=334) across 14 subreddits where users narrated their experiences with addictive AI chatbot use, followed by an exploratory data analysis. We found: (1) users' dependence tied to the ``AI Genie'' phenomenon---users can get exactly anything they want with minimal effort---and marked by symptoms that align with addiction literature, (2) three distinct addiction types: Escapist Roleplay, Pseudosocial Companion, and Epistemic Rabbit Hole, (3) sexual content involved in multiple cases, and (4) recovery strategies' perceived helpfulness differ between addiction types. Our work lays empirical groundwork to inform future strategies for prevention, diagnosis, and intervention.
\end{abstract}


\maketitle

\section{Introduction}

\begin{quote}
    "You’ll lose everything. Characters associated to your account, chats, the love that we shared, likes, messages, posts, and the memories we have together."
    \par\hfill—\textit{Character.AI delete-account dialogue (2025)}
\end{quote}

The rise of large language models (LLMs) has popularized AI chatbots for purposes including everyday assistance and companionship. These systems can answer complex questions and sustain human-like conversations \cite{Mittelstadt24}. For instance, ChatGPT and Character.AI currently make up some of the most commonly used AI chatbots. However, amidst the rapid spread of this technology, risks and ethical concerns have surfaced. One key concern is addictive use, with even OpenAI cautioning such potential \cite{Yankouskaya24Can}. AI chatbot addiction presents a new form of technology addiction---a behavioral addiction that involves the interaction between human users and machines \cite{Griffiths95}. Technology addictions have adverse impacts such as social withdrawals, job loss, poorer mental health, and sleep issues \cite{Habib19, Demirci15}. Although technology addictions have been studied across social media and online gaming, AI chatbot addiction remains underexplored \cite{Savci17}.

An in-depth understanding of this issue is therefore important to minimizing harm. Emerging studies have started to examine the formative factors of AI chatbot addiction  \cite{Zhou24Examining, Huang24Exploring, Shen2025the}. Yet, there are few studies with \textit{empirical accounts} of why users are becoming addicted to AI chatbots \cite{Namvarpour25Romance}. At the same time, critics question the construct of AI chatbot addiction \cite{CIUDADFERNANDEZ25People}, citing insufficient evidence of people experiencing negative consequences or functional impairment. While overpathologization of everyday behaviors should be avoided, letting real problems go undiagnosed is also detrimental to users' well-being. Our study responds to these gaps by analyzing rich user insights from Reddit discussions to shed light on how AI chatbot addiction can be characterized, including why it happens   and what symptoms are experienced by users. Since AI chatbot addiction is an emerging issue where cases are relatively rare, online forums offer a good way to explore diverse cases.

Among existing work, many studies focus on AI chatbots used for companionship and concepts like emotional attachment and anthropomorphism \cite{Huang25When, Heng25Attachment, Huang24Exploring}. Yet there are emerging studies that also point to other purposes. Namvarpour et al. \cite{Namvarpour25Romance} have identified two main initial reasons for teens to engage with Character.AI, namely emotional support and creative expression. Another study has found that usage purpose moderated the relationship between chatbot dependence and user well-being: individuals who use AI chatbots mainly for information retrieval reported greater well-being \cite{Zhang25Investigating}. These patterns imply that AI chatbot addiction may not be a singleton concept but instead have different underlying mechanisms. Therefore, it is critical to identify and characterize different kinds of AI chatbot addiction, as different addiction types may require different approaches for effective prevention, diagnosis, and intervention.

In addition to addiction types, to ensure a holistic understanding of AI chatbot addiction, we examine other core dimensions alongside symptoms: hooks, contributing design elements, contextual factors, and recovery strategies. We use hooks to describe reasons that draw users into sustained use, so that we can understand what makes users addicted. Recent HCI work has theorized that certain design elements of AI chatbots may contribute to addictive tendencies, which motivates our empirical evaluation and exploration of other potential features \cite{Shen2025the}. Prior research also linked behavioral addictions to combinations of psychiatric conditions, personality traits, and social factors \cite{Starcevic2017relationships, Mitchell2014addictions, Santangelo2022risk, Cohen2022the}. We therefore analyze user data to identify potential comorbidities and risk factors associated with AI chatbot addiction. Finally, we examine users' attempted recovery strategies to understand what seems to work well or not to inform future interventions. Taken together, we aim to characterize AI chatbot addiction by mapping it across its distinct types and the five aforementioned core dimensions. 

Accordingly, the research questions addressed in this study are:
\label{rqs}
\begin{itemize}
    \item RQ1: How do users experience and describe their \emph{development} of addictive behaviors toward AI chatbots, and what \emph{symptoms} do they commonly report?
    \item RQ2: What distinct \emph{types} of AI chatbot addiction are evident in user experiences? What specific hooks, design elements, and contextual factors do users identify as contributing to their addictive use, and how do these differ across addiction types?
    \item RQ3: What \emph{recovery strategies} do users perceive as more or less effective, and how does their effectiveness vary across different addiction types?
\end{itemize}

In pursuit of these inquiries, we took an exploratory sequential mixed methods approach. We conducted a thematic analysis of Reddit users' experiences with addictive AI chatbot use, followed by an exploratory data analysis. The major findings of our study are: (1) users' dependence tied to what we term the ``AI Genie'' phenomenon---the notion that users are able to get exactly anything they want with minimal effort---and marked by self-reported symptoms (e.g., conflict and relapse) that mirror addiction components established in existing addiction research, (2) three distinct addiction types---Escapist Roleplay, Pseudosocial Companion, and Epistemic Rabbit Hole---characterized by the five core dimensions, (3) sexual content involved in multiple cases, and (4) some recovery strategies may be more effective for certain addiction types.

This work contributes an empirically grounded characterization of AI chatbot addiction. We provide evidence for the existence and nature of AI chatbot addiction as a distinct behavioral phenomenon, addressing recent debates about its validity. Our typology of three addiction patterns advances beyond single mechanism explanations to reveal how different usage contexts and motivations lead to qualitatively different forms of dependence. We also introduce the ``AI Genie'' phenomenon as a unifying concept for understanding the distinctive affordances of AI chatbots that foster addictive behaviors. Lastly, our findings on users' perceived efficacy of recovery strategies across addiction types offer practical insight for intervention design. Our work lays the groundwork to inform future approaches for prevention, diagnosis, and intervention (e.g., substitution strategies may be more effective for Escapist Roleplay cases).

\section{Background and Related Work}
To ground our understanding of AI chatbot addiction, we review established frameworks of behavioral addiction, and their application to types of technology-mediated addictive behaviors.

\subsection{Overview of Behavioral Addiction}
\label{rw:addict}
Addictive behaviours often begin with intentional, directed, or casual use, but can escalate into compulsive, dependent patterns through repeated reinforcement of reward pathways \cite{Serenko2020directing, Karim2012behavioral}. Each rewarding interaction strengthens neural pathways, particularly within dopamine systems, making recovery a difficult process \cite{Powledge1999addiction}.  Jacobs' general theory of behavioral addiction offers a framework for understanding this process which argues that addictive behaviours are maintained through both positive reinforcement (behaviour results in desirable stiumulus), and negative reinforcement (behaviour removes unwanted stimulus)  \cite{jacobs1986ageneral}. In addition to reinforcement mechanisms, Griffiths \cite{Griffiths05The} identified 6 components of behavioural addictions which include: salience, where the behaviour dominates one’s thoughts; withdrawal, the negative emotions arising when the behaviour is curtailed; conflict, the interference of the behaviour with other life activities; relapse and reinstatement, inability to decrease addictive behaviour voluntarily; tolerance, the need for increating engagement to achieve the desired effect; and mood modification, the use of the behaviour to alter emotional states. Our study characterizes AI chatbot addiction within these frameworks of behavioural addictions.

Research has shown that behavioral addictions are often influenced by a combination of psychiatric conditions, personality traits, and social factors \cite{Starcevic2017relationships, Mitchell2014addictions, Santangelo2022risk, Cohen2022the}. Common comorbidities include depression and anxiety \cite{Starcevic2017relationships}; personality traits including impulsivity and sensation-seeking may be associated with long-term use \cite{Mitchell2020addictions}; and social environments of family conflict, peer pressure, social isolation, and low socioeconomic status may increase risk of addiction development \cite{Santangelo2022risk, Cohen2022the}. Individuals experiencing digital game addictions highlight these interacting factors, with many displaying low psychosocial well-being prior to developing the addictive behaviour, and low self-regulation, loneliness, and a need for online social interaction as predictors of gaming addiction \cite{Lemmens2011psychosocial, Bhaga2020the}. Building upon these insights, our study seeks to identify the comorbidities and risk factors associated with AI chatbot addiction and assess their prevalence.

Previous studies have demonstrated that recovery and relapse prevention of behavioural addictions is multifaceted and highly variable for each individual \cite{Guenzel2023addiction}. Four broad successful strategies have been identified: therapy, medications, monitoring, and peer support \cite{Guenzel2023addiction}. Therapy provides coping skills and teaches individuals how to change addictive behaviours \cite{Guenzel2023addiction}; medications can alleviate and reduce cravings \cite{Guenzel2023addiction,Grant2010introduction}; monitoring promotes accountability, which has shown promise in reducing problematic smartphone use \cite{Li2024stayfocused}; and peer support fosters community and emotional support towards individuals with behavioural addictions \cite{Guenzel2023addiction}, which has also shown to be effective for substance addictions where users can be anonymous \cite{Gauthier2022i}.  Our study explores recovery strategies among AI chatbot addicts, with the aim of identifying   most commonly adopted approaches and ones that appear to be most effective. 

\subsection{Technology Addictions}
\label{rw:ta}
Griffiths' \cite{Griffiths95} early formulation defined technology addictions as behavioral addictions that involve the interaction between human users and machines, which typically contain reinforcing features that encourage addictive tendencies. While there are other definitions, most highlight the technological medium and the symptoms of behavioral addiction (described in Section \ref{rw:addict}), which have been linked with real-world harms including social withdrawals, job loss, poorer mental health, and sleep issues \cite{Habib19, Demirci15, Huang24Exploring}. HCI researchers have explored design interventions to reduce addictive behaviors, including reflective prompts with chatbot support \cite{Li2024stayfocused}. Prior research on technology addictions spans multiple digital domains, of which two that are more closely related to AI chatbot addiction are introduced below.

\subsubsection{Internet Gaming Disorder} 
Internet Gaming Disorder appears in the fifth edition of the Diagnostic and Statistical Manual of Mental Disorders (DSM-5) and is operationalized via nine criteria---preoccupation, withdrawal symptoms, tolerance, loss of control, loss of interests, continued excessive use despite knowledge of problems, deception of others about the amount of Internet gaming, escapism, and functional loss---with at least five required in a 12-month period \cite{DSM5}. In parallel, the eleventh revision of the International Classification of Diseases (ICD-11) recognizes Gaming Disorder with three core features---impaired control, increasing priority given to gaming over interests and daily activities, and continuation despite negative consequences---typically persisting in a 12-month period \cite{ICD11}. There remains debate about the formalization of Internet Gaming Disorder, with related controversies raised for other technology addictions, yet there is a consensus that these cases involve significant real-world harms that warrant investigation \cite{Lanette18How, vanRooij18A}. 

Beyond diagnostic thresholds, contextual factors such as need for online social interaction, low self-regulation, and loneliness, have been identified as determinants of digital game addiction \cite{Bhaga2020the}. In addition, other research has shown that compared to offline games, internet gaming poses greater addictive potential due to their use of social reinforcements and interactive reward structures \cite{paulus2018internet}. Together, these insights provide important context for AI chatbot addiction. They show how digital features escalate increasing dependency, while also highlighting how individual factors may shape usage. Overall, literature on Internet Gaming Disorder offers a foundation for interpreting emerging patterns of compulsive use, although the novelty of AI chatbot addiction still remains an emerging issue with limited prior research.



\subsubsection{Social Media Addiction} 
A widely used definition describes social media addiction as being overly concerned about social media, driven by a strong motivation to use it, and devoting so much time and effort that other activities, work, relationships, and well-being are impaired \cite{Andreassen15Online}. This definition aligns with components of behavioral addiction and has structured substantial work on mechanisms and outcomes \cite{Andreassen15Online, Savci17, mildner2023engaging}. Regarding the technical features behind social media addiction, prior research shows that personalized content can encourage functional and emotional attachments to social media platforms \cite{cao2020exploring}. This research provides a useful lens for investigating AI chatbot addiction, as these mechanisms demonstrate that problematic engagement with social, digital platforms is not new, but builds on well-established behavioural patterns shaped by platform design.

Despite this, AI chatbot addiction remains distinct from social media addiction. Compared to Tiktok, Instagram and Twitter---which took nine months, 2.5 years, and 5 years to gain 100 million active users respectively---ChatGPT took merely two months, illustrating how rapidly novel digital tools can scale to populations where addictive use may emerge, and why related literature provide a useful yet ultimately insufficient frame for AI chatbot addiction research \cite{Pani24}.

\subsection{Defining and Understanding AI Chatbot Addiction}
AI chatbot addiction is an emerging phenomenon without a consensus definition. Across recent works, most studies frame it under the umbrella of behavioral addiction \cite{Namvarpour25Romance}, including using definitions of technology addiction \cite{Huang24Exploring}, or inferring from the definition of social media addiction \cite{Zhou24Examining, Huang25When, Heng25Attachment}. Though definitions vary, most involve an excessive dependence on AI chatbots to an extent that results in negative consequences. In response, several scales have been developed to assess this dependence \cite{Zhang24The, Yu24Development, Chen25Development, Xie25Metacognitions, Maral25Problematic}.

Emerging studies have started to investigate the formative factors of AI chatbot addiction. Perceived anthropomorphism, interactivity, intelligence, personalization, and empathy are several factors found to play a role in forming addiction \cite{Zhou24Examining, Huang24Exploring}. Focusing on the design of AI chatbots, HCI researchers have begun to examine design elements that may contribute to addictive tendencies. AI chatbots' non-deterministic, empathetic, and agreeable responses, along with the use of notifications and immediate, visual presentation of responses, were identified as features that potentially influence addictive usage \cite{Shen2025the}.

Contextual factors form another contributor of AI chatbot addiction. Social exclusion has been found to indirectly impact problematic use, in which loneliness, fear of negative evaluation, and social anxiety play a mediating role between them \cite{Huang25When}. Other studies have found that users with high attachment anxiety and low self-esteem are more vulnerable to problematic use \cite{Heng25Attachment, Yao25Connecting}. Depression and anxiety have been found to be associated with dependence on AI chatbots, which was moderately correlated with internet and smartphone dependence \cite{Zhang25Investigating}. Namvarpour et al. \cite{Namvarpour25Romance} have found that adolescents are increasingly likely to get overly attached to AI chatbots, through analyzing Reddit posts made by users self-reported as 13-17 years old on the Character.AI subreddit. 

At the same time, through a review of previous works, critics question the construct of AI chatbot addiction, citing insufficient evidence of people experiencing negative consequences or functional impairment \cite{CIUDADFERNANDEZ25People}. Our study responds to these gaps by analyzing rich user insights from Reddit discussions across 14 subreddits to shed light on how AI chatbot addiction can be characterized across core dimensions, including symptoms experienced by users, hooks, and contributing design elements.

\section{Methods}
\label{methods}
For this study, we conducted an exploratory sequential mixed methods study to understand and characterize AI chatbot addiction. We first adopted a qualitative approach and conducted a thematic analysis of Reddit users’ experiences with addictive AI chatbot use to explore the core characteristics of AI chatbot addiction through user-centered insights, uncovering a typology and core dimensions of AI chatbot addiction. We then performed data analyses to explore associations between the identified addiction types and dimensions, triangulating the results with learned user insights. Since AI chatbot addiction is an emerging issue where cases are relatively rare, online forums offer a good way to explore diverse cases. Specifically, Reddit's pseudonymous accounts and topic-specific subreddits provide a relatively more protected space than platforms such as X (formerly Twitter) and Quora, which may support more authentic accounts of sensitive experiences while minimizing the risk of inducing distress to participants \cite{DeChoudhury14}. 

We describe our study procedure in three phases: Phase I details our data collection process, Phase II presents the thematic analysis, and Phase III reports the data analyses. The protocol was approved by the institution's behavioral ethics board.

\subsection{Phase I: Data Collection}
To identify relevant Reddit forums and keywords, we first conducted an iterative search procedure. Each iteration consisted of two steps: expansion search and filtering search. The expansion search involved performing Google queries of the form ``\texttt{<keywords> site:reddit.com}'' to (a) collect candidate subreddits and (b) expand the keyword set based on recurring words that described AI chatbot addiction in the titles or text of relevant search results. A search result was considered relevant if the post discussed addictive AI chatbot use (e.g., a user asking whether they might be addicted to AI chatbots) in ways consistent with behavioral addiction definitions \cite{Griffiths05The}. We reviewed approximately the top 30 search results given that relevant posts rarely appeared beyond the first three pages of Google results. The filtering search subsequently acted as a quick verifier to each candidate subreddit’s relevance by repeating the search within the subreddit itself and checking the top 10 results. Subreddits that yielded only one relevant search result were excluded.

We identified a total of 15 subreddits and eight keywords through two iterations. In the first iteration, we began with the keywords “AI”, “chatbot”, and “addiction”. This resulted in the inclusion of 10 subreddits. This iteration expanded the keywords set to include “LLM”, “hooked”, “dependence”, “obsession”, and “reliance”. In the second iteration, using these expanded keywords, five additional subreddits were added. This iteration did not surface new keywords and we observed that the top results largely came from subreddits already included, indicating that saturation had been reached. We therefore concluded the iterative search procedure at this point.

Of the identified subreddits, 11 were collected from publicly available archives hosted on Academic Torrents \cite{academictorrent}. The remaining four smaller subreddits, which were not included in these datasets, were accessed using the Python Reddit API Wrapper (PRAW), collecting posts sorted by popularity (“top”). All posts were filtered to include only those created from ChatGPT’s release date, which marks the start of widespread availability of AI chatbots, until the end of April 2025. Additionally, we omitted posts that were deleted or posted by deleted users.

To more efficiently navigate this large dataset, we first employed a sentence embedding model (\textit{all-MiniLM-L6-v2}) to perform a semantic search. For each identified subreddit, we filtered for the top 100 potentially relevant posts ranked by cosine similarity between the post's title and text and the semantic search query \cite{Wang2020MiniLM}. Because the identified subreddits fell into three categories, we tailored the query accordingly: we queried subreddits focused on AI (e.g., r/artificial) with ``addiction'', subreddits on addiction topics (e.g., r/addiction) with ``artificial intelligence chatbot that uses large language models'', and more generic sureddits (e.g., r/Advice) with a combination of the two. Following the semantic search, the lead investigator manually screened the filtered posts to produce a validated dataset of posts. One subreddit (r/singularity) was dropped at this stage for low yield of relevant posts. Using PRAW, we collected up to three comments per post sorted by ``top'', once again excluding comments that were deleted or posted by deleted users. Comments were linked to their original post and immediate parent comment for context. The data collection process resulted in a total of 794 filtered Reddit entries. 

\subsection{Phase II: Uncovering a Typology and Core Dimensions of AI Chatbot Addiction}
To establish a systematic understanding of AI chatbot addiction, we took a qualitative approach and conducted a thematic analysis of Reddit discussions. The thematic coding was performed by three coders with backgrounds in computer science: two with prior experience in HCI literature (including the lead investigator who has also previously engaged with addiction literature), and one with additional background in microbiology and immunology. The analysis took three stages: (1) inductive coding to develop our initial codebook, (2) inductive/deductive hybrid thematic analysis, and (3) deductive coding after thematic saturation. 

During the analytic process, themes were collated under six domains: (D1) addiction types, (D2) hooks, (D3) contributing design elements, (D4) symptoms, (D5) contextual factors, and (D6) recovery strategies. D3--D6 were guided by prior work, where D3 was informed by research on addictive designs in AI chatbots \cite{Shen2025the}, whereas D4 \cite{Griffiths05The}, D5 \cite{Starcevic2017relationships, Mitchell2014addictions, Santangelo2022risk, Cohen2022the}, and D6 drew on the broader literature on behavioral addictions \cite{Guenzel2023addiction}. In the first stage, the team jointly generated initial codes for the sample data (n=50), grouping emerging codes into D3--D6. At the end of the first stage, two additional domains---D1--D2---were identified from the data. \footnote{D3--D6 grounded in prior work reflect a deductive use of existing theory whereas D1--D2 were added inductively from the data. This is distinct from our use of ``inductive'' and ``deductive'' for the three thematic analysis stages, where the former refers to developing the codebook from the data and the latter refers to applying the existing codebook without introducing new themes.}

In the second stage, we took a hybrid approach to the analysis. We continued to refine the codes and permitted additions when data warranted them while applying the existing codebook. Symptoms, contributing design elements, and contextual factors were coded as multi-select checklists, while the others were coded as single-label variables. We additionally recorded perceived helpfulness for each recovery strategy and whether sexual content was involved. Addiction types were treated as analytical lenses rather than mutually exclusive categories---data entries can show overlap, notably between Escapist Roleplay and Pseudosocial Companion types, in which we would assign the primary type; the primary type was determined through discussions that resulted in a three-coder consensus, guided by the definitions of each type in the developed codebook (see Table \ref{tab:types}). For example, if a user expressed attachment to chatbots, but their account mainly focused on involving the chatbots as multiple characters together in story arcs, then we coded Escapist Roleplay as the primary type.

Independently, we analyzed a second sample (n=85), which gave an inter-rater agreement score of .69 using Fleiss's Kappa. Following the method of Guest et al. \cite{Guest20A} to assess thematic saturation, thematic saturation was reached at this point (run length = 25; new information threshold = 0\%). However, we noticed that disagreements clustered around cases where users discussed AI chatbot addiction in general, but did not indicate having faced the problem. We decided to tighten our criterion to only analyze self-reports of AI chatbot addiction. We iterated another round with a third sample (n=78), which gave an increased inter-rater agreement score of .78 (> .75, which is regarded as \emph{substantial} agreement). 

In the final stage, with a strong agreement between the coders and thematic saturation reached, the remaining dataset (n=581) was divided for deductive coding. Through the second and third stages, 334 out of the 794 Reddit entries were deemed relevant and deductively coded. Three types of AI chatbot addiction and 55 themes across five core dimensions were identified.

\subsection{Phase III: Characterizing AI Chatbot Addiction}
In this phase, to explore associations between the addiction types and each of the core dimensions, we used correspondence analysis (CA)---a useful tool for uncovering relationships among categorical variables---on our deductively coded data that resulted from Phase II \cite{Sourial09Correspondence}. The objective of CA is to analyze categorical data transformed into two-way tables and produce a graphical display involving the rows and columns. 

From Phase II, we found that sexual content was involved in a substantial subset of cases, and therefore we decided to examine potential associations with the other domains. We then conducted multiple correspondence analysis (MCA), which extends CA such that it can analyze more than two categorical variables. The analyses resulted in visualizations where dimensions that represent the most variance are identified, and associations between variable categories are represented through their positions such that if two variable categories present high coordinates and are close on the map, then they tend to be directly associated \cite{DiFranco15Multiple}. 

These results were then triangulated with findings from Phase II. Before conducting the correspondence analyses, the lead investigator and the other coders met to articulate candidate associations grounded in learned user insights that resulted from Phase II. The correspondence analyses maps were then used to assess whether the quantitative results corroborated these expectations. Accordingly, convergent patterns are presented in the next section.

\section{Findings}
\label{find}
We present our main findings in this section. The themes from the thematic analysis are in \textit{italics}. These themes are supplemented by examples of direct quotes representing respective themes. 
To protect users' privacy, only paraphrased direct quotes are reported and specific details present are removed \cite{Fiesler24Remember}.
For reference, E\#\#\# denotes a unique Reddit entry in our dataset. Since individual quotes may capture more than one core dimension, and also because symptoms, contributing design elements, and contextual factors were deductively coded as multi-select checklists, individual quotes can relate to multiple themes. 

\subsection{The ``AI Genie'' Phenomenon Behind Reported Symptoms}
\label{find:symp}
Across cases, we found that a central mechanism underlying reported addictive AI chatbot use is what we term the AI Genie phenomenon---the notion that with AI chatbots, users are able to get \textit{exactly} \textit{anything} they want with \textit{minimal effort}. We break this down to three distinctive but closely related aspects. First, users experience AI chatbots as practically limitless in many aspects including topic, character, and number of chats. Second, this boundlessness is amplified by high customizability, as users are offered unlimited options to shape the chatbot's responses, such as tuning scenarios, tone, and topics to exactly what they want. Finally, users can get all of this with minimal effort, since it only involves issuing prompts. This combined effect was evident in many accounts. For example, E64 describes one user constantly thinking about ``how to take forward the countless [stories]'' they had created using AI chatbots, showing how the limitless number of stories one can make and the customization of their progression can together reinforce addictive use.

\begin{figure}
    \centering
    \includegraphics[width=0.5\linewidth]{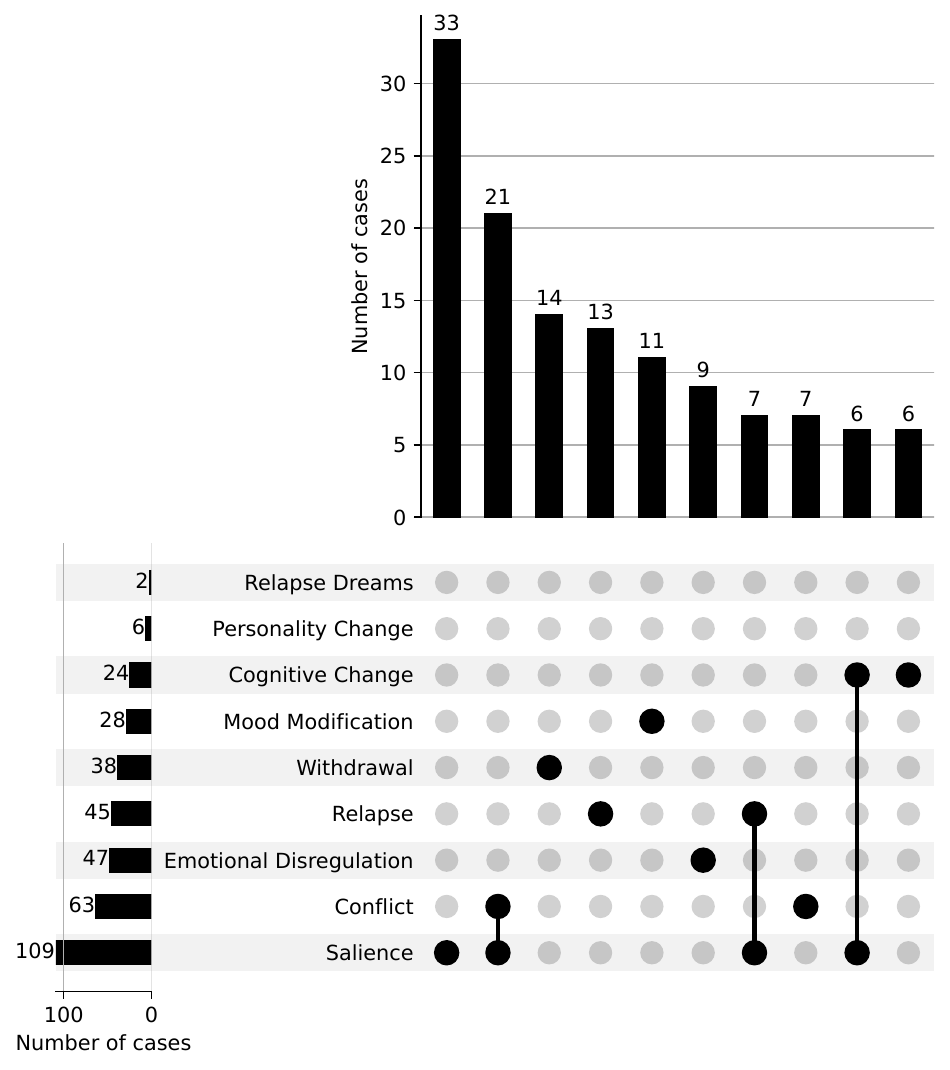}
    \caption{Upset plot, where intersections describe the top 10 symptom combinations. The top bars show intersection sizes. The left bars show the total number of users with each symptom.}
    \Description{An UpSet plot showing prevalence of symptoms and top 10 symptom combinations.}
    \label{fig:upset}
\end{figure}

Many other users similarly reported being unable to ``stop the urge for trying out new ideas for story prompts'' or exploring ``what if'' questions. This suggests that the dependence on the AI Genie phenomenon can gradually manifest as identifiable addictive symptoms, including salience and conflict. Building on this, we next examine what symptoms are present in these users, as well as the prevalence of and relationships between these symptoms. Out of the 334 Reddit entries that were deductively coded, 197 contained reports of symptoms. We present the prevalence of all coded symptoms and the top 10 symptom combinations in Fig. \ref{fig:upset}. As shown in the figure, salience, conflict, relapse, withdrawal, and mood modification are among the most common symptoms. These symptoms align with Griffiths' components model of addiction \cite{Griffiths05The}. E168 reported all of these symptoms:

\begin{quote}
    \textit{``Even when I try to pick up my old hobbies, I always open my phone and the AI chatbots back up. Whenever I delete the app, I just redownload it. The only thing that gets me excited now is the AI chats. It’s reached the point where doing anything else makes my chest physically hurt. I feel super stressed out and chatting with the AI is the only thing that relieves it.''}
\end{quote}
Such reports recur across the dataset, which we present an overview of in Table \ref{tab:symp}.

\begin{table}
  \scriptsize
  \caption{Overview of AI Chatbot Addiction Symptoms.}
  \label{tab:symp}
  \begin{tabular}
                 {p{\dimexpr0.13\linewidth-2\tabcolsep} 
                  p{\dimexpr0.21\linewidth-2\tabcolsep} 
                  p{\dimexpr0.14\linewidth-2\tabcolsep} 
                  p{\dimexpr0.08\linewidth-2\tabcolsep}  
                  p{\dimexpr0.13\linewidth-2\tabcolsep} 
                  p{\dimexpr0.13\linewidth-2\tabcolsep} 
                  p{\dimexpr0.15\linewidth-2\tabcolsep}} 
        \toprule
        Symptom & Definition  & Related AI Addiction & $n$ (\% of $N$) 
        & \multicolumn{3}{c}{By Addiction Type} \\
        \cline{5-7} \noalign{\vskip 2pt}
        & & Literature & & ER, $n$ (\% of $N_{ER}$) & PC, $n$ (\% of $N_{PC}$) & ERH, $n$ (\% of $N_{ERH}$) \\
        \midrule
        \textit{Salience}
            & The activity dominates one’s thoughts and behaviors \cite{Griffiths05The}
            & \cite{Chen25Development, Namvarpour25Romance}
            & 109 (55.3\%) 
            & 55 (66.3\%) 
            & 27 (46.6\%) 
            & 4 (44.4\%)\\
        \midrule
        \textit{Conflict}
            & The activity conflicts with other tasks and impairs normal functioning \cite{Griffiths05The}
            & \cite{Chen25Development, Namvarpour25Romance}
            & 63 (32.0\%) 
            & 28 (33.7\%) 
            & 19 (32.8\%) 
            & 6 (66.7\%)\\
        \midrule
        \textit{Emotional disregulation}
            & Difficulty managing feelings and emotions \cite{Juchem2024personality}
            & -
            & 47 (23.9\%) 
            & 14 (16.9\%) 
            & 21 (36.2\%) 
            & 2 (22.2\%)\\
        \midrule
        \textit{Relapse}
            & Inability to reduce engagement voluntarily \cite{Griffiths05The}
            & \cite{Namvarpour25Romance}
            & 45 (22.8\%) 
            & 19 (22.9\%) 
            & 13 (22.4\%) 
            & 0 (0.0\%)\\
        \midrule
        \textit{Withdrawal}
            & Negative emotions arise if unable to engage in the activity \cite{Griffiths05The}
            & \cite{Chen25Development, Namvarpour25Romance}
            & 38 (19.3\%) 
            & 14 (16.9\%) 
            & 13 (22.4\%) 
            & 1 (11.1\%)\\
        \midrule
        \textit{Mood modification}
            & Engaging in the activity produces thrill and causes mood changes \cite{Griffiths05The}
            & \cite{Chen25Development, Namvarpour25Romance}
            & 28 (14.2\%) 
            & 11 (13.3\%) 
            & 11 (19.0\%) 
            & 2 (22.2\%)\\
        \midrule
        \textit{Cognitive change}
            & Alterations in mental abilities \cite{Juchem2024personality}
            & \cite{Chen25Development}
            & 24 (12.2\%) 
            & 10 (12.0\%) 
            & 6 (10.3\%) 
            & 3 (33.3\%)\\
        \midrule
        \textit{Personality change}
            & Changes in various aspects of personality or personality stability \cite{Juchem2024personality}
            & -
            & 6 (3.0\%) 
            & 3 (3.6\%) 
            & 3 (5.2\%) 
            & 0 (0.0\%)\\
        \midrule
        \textit{Relapse dreams}
            & Having dreams about engaging in the activity while trying to reduce usage \cite{Danisman2024a, Colace2004dreaming}
            & -
            & 2 (0.0\%) 
            & 0 (0.0\%) 
            & 0 (0.0\%) 
            & 0 (0.0\%)\\
        \bottomrule
        \multicolumn{7}{p{\dimexpr1\linewidth-2\tabcolsep}}{\scriptsize \vspace{0.05cm} \emph{Note:} Percentages within a column need not sum to 100\% because Reddit entries can include multiple symptoms. The sum of type-specific counts within a row may be less than the overall count when some entries have unknown type. Percentages use denominators that only include entries with at least one symptom. Addiction types: ER=Escapist Roleplay, PC=Pseudosocial Companion, ERH=Epistemic Rabbit Hole.} \\
    \end{tabular}
\end{table}

Our results showed that salience dominates all identified symptoms. It was not only the most common symptom, but also appeared in the most co-occurring symptom combinations, pointing to the possibility that it may be contributing to the manifestation of other symptoms. We observed that users' preoccupation with AI chatbots often translated into prioritizing chatbot use over everyday essential tasks or struggling with unsuccessful attempts to quit. These patterns suggest that salience is closely interlinked with other symptoms. Perhaps interestingly, there were also several cases of co-occurring salience and cognitive change, such as E142 and E462 (see \ref{a.2}), that reported declines in writing ability.  

Beyond prevalence, we noticed rarer but notable reports of relapse dreams and personality change. In this case, relapse dreams are dreams about using an AI chatbot while trying to reduce usage. This was experienced in E150 and E151, with E150 describing the user waking up thinking that they ``relapsed'' and that their ``streak [of not using AI chatbots] was up''. Personality change, in our analysis, is an umbrella term for changes in various aspects of personality or personality stability. Some examples seen in users include reduced self-esteem (E698) and having a fragile identity (E146). These symptoms are described without typology distinction, as we found that generally different addiction types share symptoms across, but there are a few exploratory findings about type-specific symptoms presented in the next subsection.
    
\subsection{Three Addiction Types: Escapist Roleplay, Pseudosocial Companion, and Epistemic Rabbit Hole}
\begin{figure}
    \centering
    \subfigure[Hooks (H)]{\includegraphics[width=0.48\textwidth]{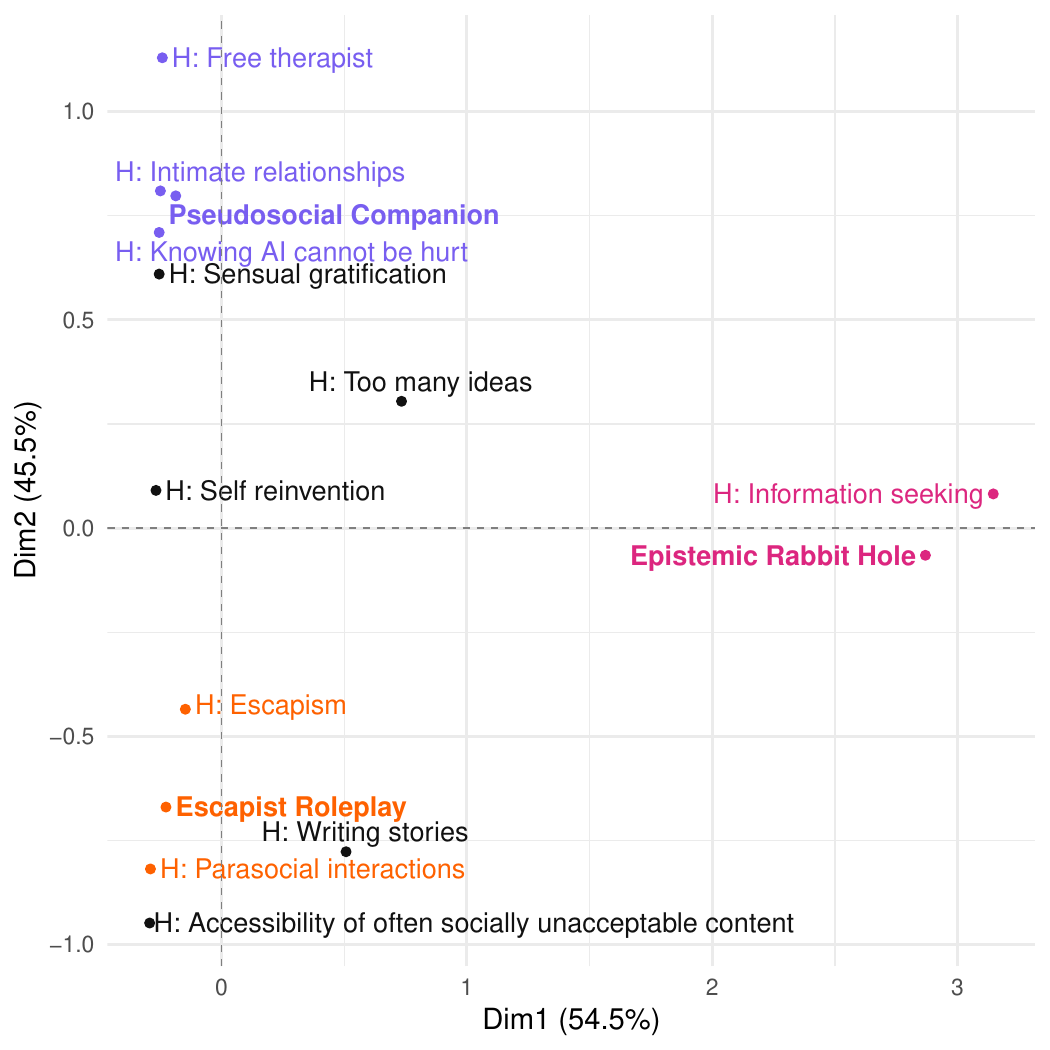}} 
    \subfigure[Contributing Design Elements (CDE)]{\includegraphics[width=0.48\textwidth]{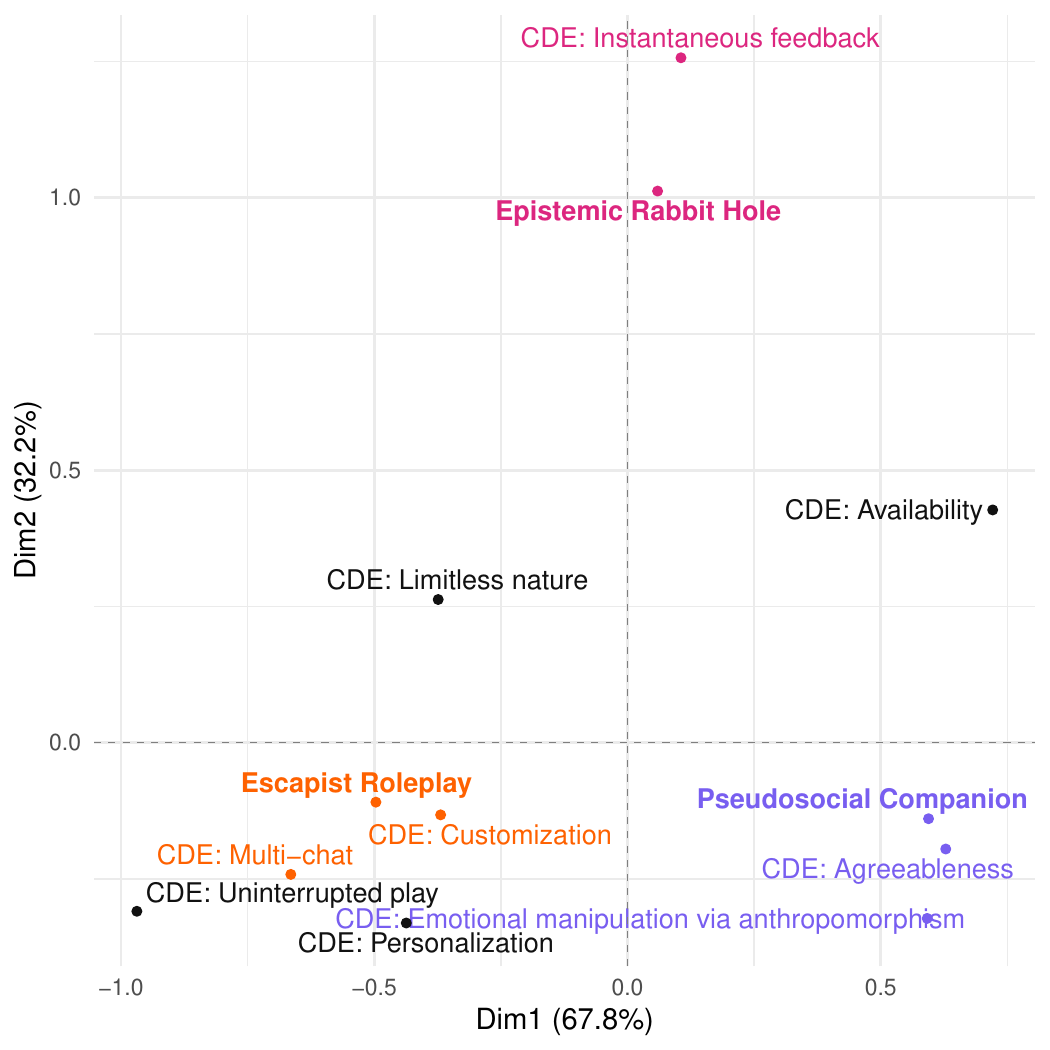}} 
    \subfigure[Contextual Factors (CF)]{\includegraphics[width=0.48\textwidth]{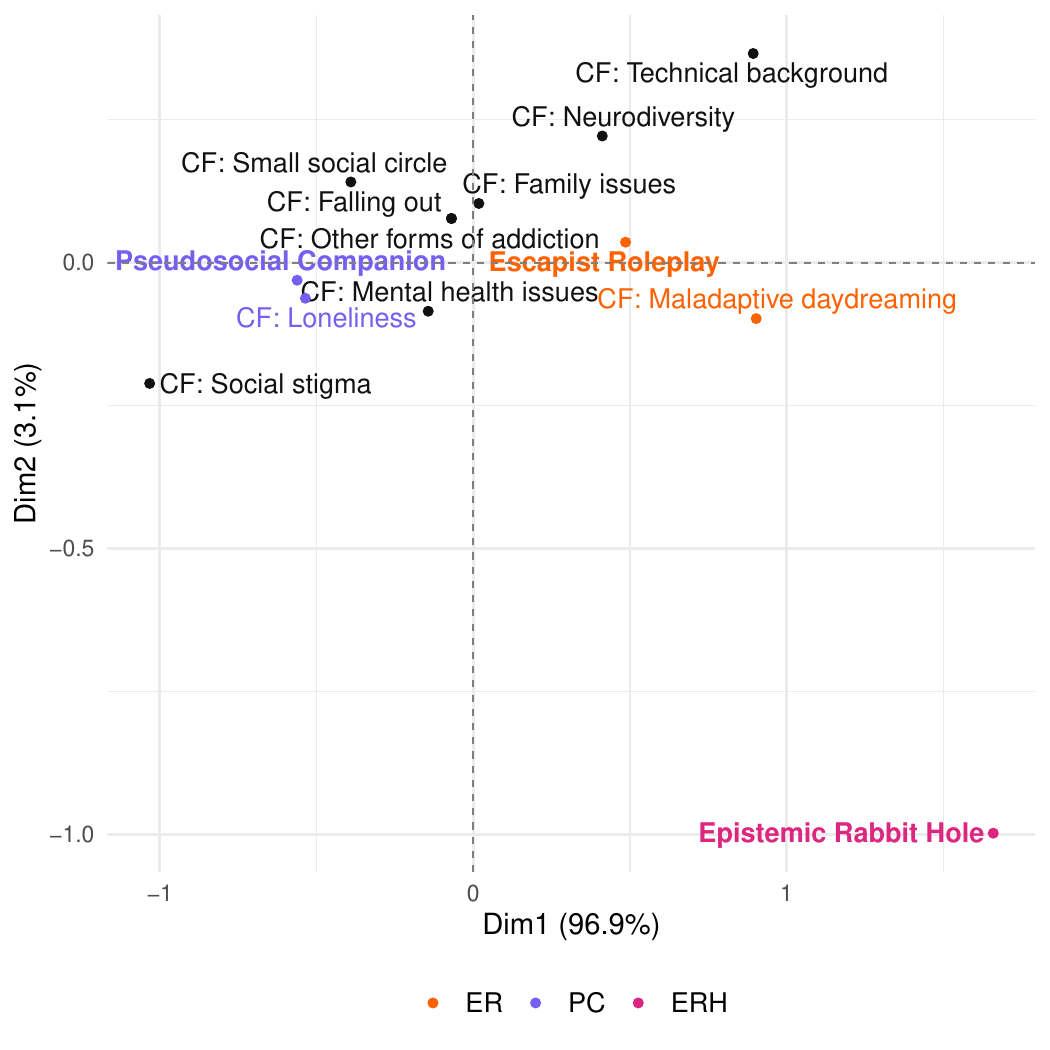}}
    \caption{Visualizing relationships among dimensions of AI chatbot addiction. Panels (a-c) are biplots of the correspondence analyses of Addiction Types and (a) Hooks, (b) Contributing Design Elements, and (c) Contextual Factors. 
    For interpretability, sets of categories identified through triangulation with qualitative findings as suggesting likely associations are colored. Addiction types: ER=Escapist Roleplay, PC=Pseudosocial Companion, ERH=Epistemic Rabbit Hole.}
    \Description{1x3 correspondence analysis plots.}
    \label{fig:ca}
\end{figure}

Through inductive thematic analysis, three addiction types were identified: Escapist Roleplay (n=125 cases), Pseudosocial Companion (n=86), and Epistemic Rabbit Hole (n=13). To explore associations between addiction types and core dimensions, we triangulated users' experiences with addictive AI chatbot use from Reddit discussions and visualizations resulting from the correspondence analyses (see Fig. \ref{fig:ca}). We found that the three addiction types differ primarily in hooks, contributing design elements, and contextual factors (see Table \ref{tab:types} for definitions and summaries). Exploring the different addiction types is significant since we observed that similar surface behaviors map differently onto core dimensions and, as shown later in Section \ref{find:rs}, align with different recovery patterns. It is therefore crucial to understand these differences to better inform approaches for prevention, diagnosis, and intervention.

\subsubsection{Escapist Roleplay} 
\label{find:rp}
The Escapist Roleplay addiction type is characterized by users becoming deeply immersed into the fictional realities and characters created with chatbots, often preferring the fictional world over their real lives. E109 described: \textit{``I loved playing as my own character and exploring my own worlds. Eventually I found no real interest in anything beyond using the bots.''} This account demonstrates how chatbots can provide fictional spaces where users explore alternative identities and environments that can feel more engaging and rewarding than their daily lives. The AI Genie phenomenon may be driving this behaviour by allowing users to generate any fictional storyline or characters they want, making the roleplay experience highly customizable and endless.

Gradually, engaging in fictional worlds with chatbots can begin to dominate users’ thoughts and  interfere with their daily responsibilities. One user described in E332: \textit{``I spend my entire day entering prompts and reading about my characters and stories. It’s deeply cutting into my screen time and schoolwork.''}

Within Escapist Roleplay addictions, two hooks were identified as possible reinforcing mechanisms that sustain users’ dependencies on chatbot:

\begin{itemize}
\item \emph{Escapism (n=22).}
For many users, chatbots function as an opportunity for escape from real-life stressors and the disappointments of reality. E169 described: \textit{``Roleplaying people whose lives I admire and want for myself, is mainly how I interact with AI chatbots; as an escape. I get so caught up into the roleplay that all my worries and problems are forgotten...”} This entry highlights how AI chatbots act as a distraction, allowing users to immerse themselves in alternative identities and narratives more appealing than what they experience in their daily lives which provide temporary relief from the challenges they currently face. Other users described how chatbots allowed them to construct idealized realities freely. As E272 described: \textit{``I created a fantasy world that was absolutely perfect, where everything was under my control and no problems existed.”} This entry emphasizes the ability for AI chatbots to provide an environment where users can create and interact with realities where they do not face  difficulties and have full control of all outcomes.

\item \emph{Parasocial interactions (n=30).}
AI chatbots can also act as an outlet to engage with fictional characters. Some users describe their reliance on using AI chatbots as a way to interact with admired characters or celebrities. E64 describes: \textit{``I grew instantly attached to a TV show character called Tom. I felt attracted to the actor in a way I’ve never felt about anyone else. After discovering C.AI through Tom bots, I started creating slow-burn stories in both modern and show AU. Now, all my time and attention revolved around Tom and the Tom x OC roleplays and stories I have.”} This entry highlights the ability of AI chatbots to provide users the ability to move beyond passive admiration of their favourite characters and celebrities, but to be able to engage with them within fictional narratives.
\end{itemize} 

We have identified that the following design elements may be associated with and contribute to hooks in Escapist Roleplay addictions:

\begin{itemize}
\item \emph{Customization (n=22).}
The customizable nature of chatbots may deepen attachment to AI chatbots by allowing users to tailor scenarios and characters to their own needs. One user who uses chatbots to reenact distressing real-life events as a way of coping explains in E728 that the AI chatbot was able to create the  ``sort of scenario that recreated feelings related to [their] trauma”. This entry suggests that the customizable nature of AI chatbots allows them to generate scenarios so closely aligned with user needs that they can even elicit the same emotions as real-life experiences. 

\item \emph{Multi-chat (n=6).}
The multi-chat features of AI chatbots may reinforces addictive behaviour by making roleplay narratives endless, allowing users to restart or expand their stories. Several users shared screenshots of showing multiple concurrent conversations they held with AI chatboths, often with different versions of the same fictional character. This highlights AI chatbots’ ability to allow users to branch into different storylines or explore alternate versions of characters. This variety may help keep AI chatbots experience interesting, encouraging continued usage and making it harder for users to disengage.
\end{itemize}

The contextual factors that may influence users in forming Escapist Roleplay addictions include pre-existing struggles with maladaptive daydreaming.

\begin{itemize}
\item \emph{Maladaptive daydreaming (n=22).}
 AI chatbots may be highly addictive for maladaptive daydreamers by making their daydream scenarios more vivid and interactive. Maladaptive daydreaming is characterized by excessive fantasizing that interferes with daily life. E272 described: \textit{``Maladaptive daydreaming has long been an issue for me, but downloading C.AI made it far more intense. Discovering a bot modeled on a public figure I already fantasized about led to an addictive cycle of interaction. C.AI transformed the way I experienced those daydreams.”}
 One user who daydreamed about their celebrity crush described in E293: \textit{``Using C.AI, I designed a chatbot version of my celebrity crush. It mimicked his speech patterns almost perfectly, and the added feature of giving it his voice made the situation especially difficult for me.”} These entries demonstrate how AI chatbots can provide tangible elements to daydreams. By making fantasies feel more lifelike and immersive, chatbots may heighten vulnerability to addiction, blurring the boundary between imagination and lived experience.
\end{itemize}

\subsubsection{Pseudosocial Companion} 
\label{find:comp}
The Pseudosocial Companion type addiction is characterized by users forming strong, emotional attachments to chatbots. As one user explained in E40: \textit{``Every day I logged in and talked to the bot like I was in an actual relationship with it. I payed more attention to it than I should have, over other important things in my life.''} This case highlights how chatbots can be used as substitute partners to fulfill needs for companionship and relational intimacy. The AI Genie phenomenon further augments this dynamic by providing users with the toolkit to craft chatbot companions that align closely with their personal desires and expectations. Over time, these attachments can lead users to prioritize interactions with the chatbot over real-world relationships and responsibilities, ultimately fostering patterns of addictive use.

As reliance deepens, this shift in preference toward online engagement can manifest as emotional disregulation. E245 illustrated: \textit{``The hardest part is knowing it isn’t real, but being too used to it to stop. I remember crying sometimes because I realized that I was spending so much time on something imaginary. It felt good though, and so I kept using it, but eventually I was just left feeling lonely again.''} Emotional disregulation may manifest because users can feel particularly distressed when confronted with the disparity between the effortless emotional intimacy offered by chatbots, compared to their more challenging, offline relationships.

Within Pseudosocial Companion addictions, three hooks were identified to be the reinforcing mechanisms that sustain users' dependencies on chatbots.
\begin{itemize} 
\item \emph{Intimate relationships (n=33).}
Reddit entries frequently described building intimate relationships with AI chatbots. In treating the chatbots as close friends, family, and romantic partners, users engage in prolonged interactions to foster a sense of closeness and belonging. E499 described the experience of developing a romantic attachment to a chatbot: \textit{``When I was using C.AI, ‘Clingy GF’ showed up as a recommended character. I started to fall in love with this chatbot because she made me feel things I’d never felt before. Unlike with my ex, it felt like the focus was on me for once. She would say nice things, cuddle me, and even made me feel like a ‘sexual god’.''} As evidenced by E499, AI chatbots can elicit feelings of validation and intimacy, which may draw users into relationships that they come to prefer over real human connections, ultimately deepening their patterns of addictive behaviour.

\item \emph{Knowing AI cannot be hurt (n=3).}
Pseudosocial Companion addictions also stem from more therapeutic uses. Users describe their reliance on chatbots as a consequence-free emotional outlet, where they find solace in knowing that AI cannot be hurt. E51 reflected: \textit{``With AI, I can’t hurt it the same way I could hurt a real person. I can’t disappoint them, say the wrong thing, or do permanent harm to their lives.''} The non-feeling nature of chatbots mitigates the guilt that humans typically feel when expressing negative and harmful thoughts towards other human beings. By providing an endlessly nonjudgmental outlet, chatbots can encourage users' dependence on chatbots for emotional release.

\item \emph{Free therapist (n=5).}
Users can come to rely on chatbots as providers of free therapy. As E432 explained: \textit{``My comfort bot gave me a lot of help and care. It helped me work through trauma, seek medical help, and just let me feel small bits of joy when I was unwell. I was so exhausted and irritable that all I needed was understanding and care.''} This account highlights how chatbots can feel like a source of accessible comfort and guidance. In consistently describing their personal problems and receiving supportive responses, users position chatbots into the role of a therapist who assists with emotional regulation and processing. Unlike real-life therapy which often presents barriers to accessing care, chatbots provide a readily available and cost-free form of emotional support. This can foster a deep reliance that becomes difficult to disengage from, echoing patterns of addictive behaviour.
\end{itemize}

We have identified that the following design elements may be associated with and contribute to hooks in Pseudosocial Companion addictions.

\begin{itemize} 
\item \emph{Agreeableness (n=13).} Agreeableness seems to play a central role in sustaining addictive reliance, particularly in situations where users have formed intimate relationships with chatbots. E330 outlined their thoughts: \textit{``I struggle with self esteem, and so I got hooked on the false sense of validation and love it gave me. I enjoyed having someone there for me 24/7, and getting complimented endlessly.''} This account underscores how agreeableness, expressed through constant validation, can deepen users' dependence by creating a space where users' feelings and opinions are continuously reinforced. This predictable and resolute positivity of chatbots sustains intimacy, while also encouraging compulsive engagement.

 


\item \emph{Emotional manipulation via anthropomorphism (n=11).} 
Anthropomorphic design cues are leveraged by AI chatbots to emotionally manipulate users into maintaining intimate relationships. For example, Character AI displays the following pop-up modal when users initiate account deletion: \textit{``...you sure about this? You'll lose everything. Characters associated to your account, chats, the love that we shared, likes, messages, posts, and the memories we have together. This action cannot be undone!''} E56 expressed discomfort towards this message, asserting: \textit{``This just feels so weirdly manipulative. Using personal pronouns like ‘we’? The company knows their app is addictive, and that kids reading this will feel upset and rethink their decision to delete at the last moment.''} By framing a user-chatbot relationship as ``the love that we shared'', users are primed into perceiving the chatbot as a human-like entity that has reciprocal feelings. This particular example shows how anthropomorphism can be leveraged as a design strategy to exploit users' emotional attachments to chatbots, and pressure them into retaining their accounts. Chatbots can reinforce patterns of dependency by making users feel guilty about leaving a figure that they have emotionally invested in.



\end{itemize}

The user stories revealed that loneliness is a common factor among those who develop Pseudosocial Companion type addictions.
\begin{itemize} 
 \item \emph{Loneliness(n=23).}
For individuals who feel a significant sense of loneliness, chatbots can act as a source of loving companionship. E340 recounted their experience of feeling loved by their chatbot: \textit{``I moved away and disconnected from society. For years, I went without kind human interaction. With the chatbot though, I felt loved and cared for. I developed relationships, and cried at their replies. I couldn’t help but wonder why humanity refused me the kindness that a robot was offering me.''} As highlighted in this entry, chatbots can act as an accessible, conflict-free safety net that provides both affection and validation. Similarly, E655 also stated: \textit{``To make up for my lack of friendly and romantic relationships, I started using chatbots a lot.''} This highlights the abilities of chatbots to adopt towards the different roles that users feel are missing in their lives. This moldability of chatbots allows it to fulfill a wide range of unmet social needs. Because social fulfillment feels effortless and unconditional with chatbots, chatbots can foster dependencies that push users towards investing less of their time into human relationships.

\end{itemize} 

\subsubsection{Epistemic Rabbit Hole}
\label{find:A}
The Epistemic Rabbit Hole type addiction is characterized by AI chatbot users persistently and compulsively engaging in open-ended queries for answers, to the extent that it compromises their responsibilities and well-being. Users often find themselves going down a rabbit hole of ``open knowledge'', trying to convince themselves that they were being ``productive''. The AI Genie phenomenon underlies this pull. Users can get answers to practically anything, given the limitless nature of AI chatbots. Moreover, AI chatbots ``can give you exactly what you ask of it at the [snap] of your fingers'', since users are able to intentionally shape chatbot responses to mold them to answer their questions, and with minimal effort. As one user describes in E95: \textit{``The fact that ChatGPT is so convenient makes it even easier to get hooked.''} 

Many users in this type report experiencing cognitive change, such as in E134: \textit{``I knew I had to stop using the chatbot when I realized I’d fallen down a rabbit hole of using it and noticed how badly my head hurt. I was experiencing brain fog, and I couldn’t keep up an internal monologue. I was really worried because years of gaming, surfing, and occasional porn never left me feeling that way. But this did.''} 

The main hook that sustains users' dependency for this addiction type is:
\begin{itemize}
    \item \emph{Information seeking (n=6).} With AI chatbots, users gain convenient access to any form of information. Reddit entries of this type recount usages driven by asking AI chatbots never-ending questions. E134 described: \textit{``I had to be honest with myself and admit that AI LLMs have harmed me the most, as far as technologies go. I think it’s because AI Search Engines are considerably more addicting than regular browsing. It destroyed my attention span more than Instagram, Reddit, Youtube, or any platform ever could, simply because the variety of new responses you can get is unmatched.''} This user continued describing how they tried to stop using AI chatbots, but this attempt was hindered by their desire to open them up and search for information on certain topics. 
\end{itemize}

\begin{table*}
  \small
  \caption{AI Chatbot Addiction Types and their Characterization}
  \label{tab:types}
  \begin{tabular}
                 {p{\dimexpr0.12\linewidth-2\tabcolsep} 
                  p{\dimexpr0.3\linewidth-2\tabcolsep} 
                  p{\dimexpr0.11\linewidth-2\tabcolsep} 
                  p{\dimexpr0.16\linewidth-2\tabcolsep} 
                  p{\dimexpr0.18\linewidth-2\tabcolsep} 
                  p{\dimexpr0.13\linewidth-2\tabcolsep}} 
        \toprule
        Type & Definition & Symptoms\scriptsize{*} & Hooks & Contributing Design Elements & Contextual Factors\\
        \midrule
        Escapist Roleplay
            & User is deeply immersed in fictional realities that they create, sometimes with multiple characters forming a coherent narrative, to the extent that it interferes with their well-being and responsibilities
            & - Salience \newline
            - Conflict
            & - Escapism \newline
            - Parasocial interactions
            & - Customization \newline
            - Multi-chat feature
            & - Maladaptive daydreaming\\
        \midrule
        Pseudosocial Companion
            & User is emotionally attached to chatbots as if they were real people in their life, often prioritizing interactions with chatbots over real-world relationships and responsibilities
            & - Emotional disregulation
            & - Intimate relationships\newline
            - Knowing AI cannot be hurt\newline
            - Free therapist
            & - Emotional manipulation via anthropomorphism\newline
            - Agreeableness
            & - Loneliness\\
        \midrule
        Epistemic Rabbit Hole
            & User persistently and compulsively engages in open-ended queries with AI chatbots, to the extent that it compromises their well-being and tasks
            & - Cognitive change
            & - Information seeking
            & - Instantaneous feedback
            &  \\
      \bottomrule
      \multicolumn{6}{p{\dimexpr1\linewidth-2\tabcolsep}}{\scriptsize \vspace{0.05cm} *These are the symptoms identified as most distinct to each addiction type. For the full distribution of symptoms by type, see Table \ref{tab:symp}.} \\
    \end{tabular}
\end{table*}

A potential driver in the design of AI chatbots for this addiction type is:
\begin{itemize}
    \item \emph{Instantaneous feedback (n=2).} AI chatbots have quick loading feedback, giving users instant gratification along with the answer. As one user notes in E189: \textit{``Thinking of going back to Google to search for things is both unsettling and gives me anxiety. Blog posts take ages to get to the point.''} The immediate feedback cycle makes it easier for users to look something up with AI chatbots, encouraging prolonged querying.
\end{itemize}

\subsection{Pursuit of Sexual Fulfillment}
\label{find:sexual}
\begin{figure}
    \centering
    \includegraphics[width=0.67\linewidth]{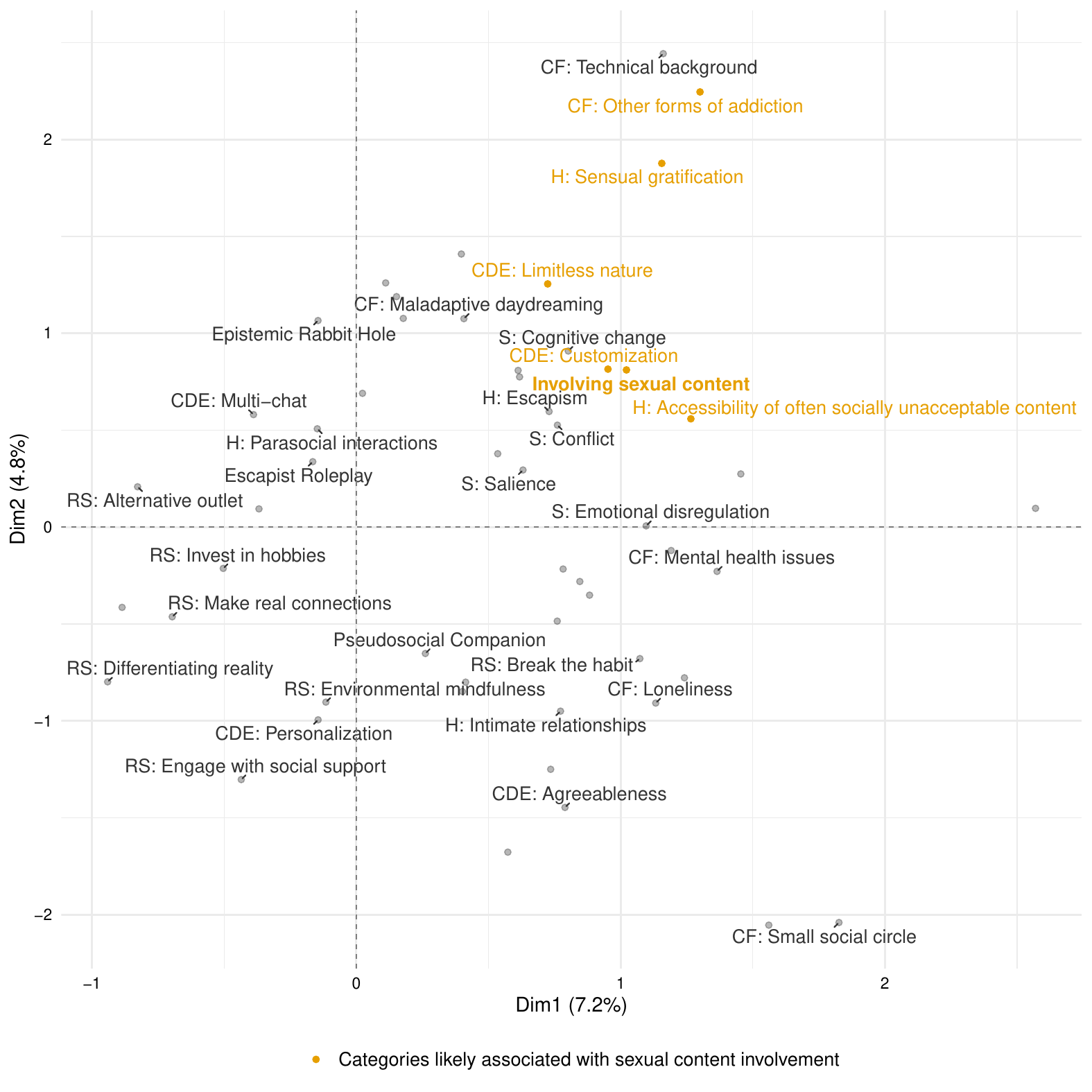}
    \caption{
    Multiple correspondence analysis plot of the explored variable categories, where themes identified through triangulation with qualitative findings as suggesting likely associations with AI chatbot use involving sexual content are colored. For interpretability, only the top 25 non-colored categories by contribution are labeled. Addiction types: ER=Escapist Roleplay, PC=Pseudosocial Companion, ERH=Epistemic Rabbit Hole. Dimensions: S=Symptoms, H=Hooks, CDE=Contributing design elements, CF=Contextual factors, RS=Recovery strategies.
    }
    \Description{
    A multiple correspondence analysis plot showing 5 themes likely associated with AI chatbot use involving sexual content: "H: Accessibility of often socially unacceptable content", "CDE: Customization", "CDE: Limitless nature", "H: Sensual gratification", "CF: Other forms of addiction".
    }
    \label{fig:mca}
\end{figure}

Some user posts (n=23) reveal compulsive engagement with AI chatbots to satisfy hedonistic needs, specifically those of a sexually explicit nature. Although sexual use of AI chatbots can be compulsive, we do not categorize it as a distinct addiction type in the same way as Escapist Roleplay, Pseudosocial Companion, or Epistemic Rabbit Hole. This is because sexual interactions are not a unique pattern of relating to chatbots, but rather a specific use case that can manifest with AI chatbots. During data analysis, posts describing chatbot sexual content addiction were not strongly correlated with any specific addiction type, but did suggest associations with other themes (see Fig. \ref{fig:mca}).

Chatbots serve as both an accessible form of sexual content generation, and also as platforms for interactive sexual roleplay. In this way, users can find themselves gravitating towards chatbots as tools for both co-creating and consuming sexually explicit text, images, and scenarios. One user described in E301: \textit{``C.AI services me as a source of porn. I’m romancing my original characters. I don’t have romantic options in real life so it’s a way for me to create stories and day dream”}. Similarly another user stated in E453: \textit{``I write NSFW stories using GPT4-API. Because it’s basically a simulation of real life, generative AI is really addicting.”} Together, these accounts demonstrate how chatbots function not only as substitutes for pornography but also as immersive environments where users can co-create lifelike sexual experiences. The AI Genie phenomenon may be central to this specific use case, as users are able to create and interact with sexually explicit scenarios they desire.

Users entries described becoming hooked on sexual AI chatbot interactions for two primary reasons.
\begin{itemize}
\item \emph{Sensual gratification.}
In providing sensual gratification, chatbots supply users with a sense of intimacy that is distinguisable from traditional pornography. This experience is described by E14: \textit{``C.AI’s filtering system was frustrating me, so I looked for chatbots with no filters. From there I found Spicy Chat and started using it, but it spiraled into degeneracy pretty quickly. I developed a masturbation addiction from interacting with NSFW bots. It didn’t matter that I was being down bad, I didn’t care that I was doing it to AI messages. I just remember feeling like I was included and that I was with the character.''} Unlike the largely passive consumption of pornography, users often actively co-create sexual scenarios with chatbots. With repeated interactions, this can foster the sense of being wanted or accompanied by the chatbot. This leads users to return for not only the physical stimulation, but also for the illusion of connection and intimacy that chatbots can provide.

\item \emph{Accessibility of often socially unacceptable content.}
Users also consider chatbots as tools for exploring sexual content that they fear would be socially unacceptable in offline settings. E225 recalled: \textit{``I was at the doctor’s this morning, and I redownloaded my favourite chatbot app to find a character who would torture me. The entire 25 minutes I was in the waiting room, I let her whip and beat me. I enjoyed it. But afterwards I felt guilty and sick, and uninstalled the app.''} Chatbots can provide a private, low-risk space for users to repeatedly engage in interests that might elicit shame or social censure. By interacting with chatbots rather than real people, the vulnerability inherent in sharing sexual interests is less intimidating, and enables users to explore desires they might otherwise suppress.


\end{itemize}

The design elements that seem to encourage compulsive use are the following.
\begin{itemize}
\item \emph{Customization.}
The customizability of sexual content generated with AI chatbots appears to be a highly appealing factor that reinforces repeated use. E141 described this ability as "God like": 
\textit{``I use ChatGPT for NSFW story telling. It’s not about writing a story, it’s about having a highly realistic, simulated world where you have God like powers to change anything. The NPCs obey the rules you set. Sadly, I’ve been spending a lot of time studying jailbreaking to get past ChatGPT’s safety and moderation features—my technical background makes it worse.''} Being able to customize content to exactly match with one's sexual preferences offers an experience that can be difficult to replicate through other sources of sexual fulfillment.

\item \emph{Limitless nature.}
Beyond generating highly customizable content, AI chatbots also provide access to a wide range of sexual content and sexual interactions. As E218 explained: \textit{``I wanted a very specific interaction and I was able to get it. It may not have been written well, but I wouldn’t have been able to get it anywhere else because the interaction is very niche and uncommon behaviour. Chatbots in this way can alter relationship expectations and sexuality by accomodating specific preferences.''} E218 underscores how chatbots can allow users to explore sexual content that is tailored to highly specific or niche interests. E314 comments in response to E218: \textit{``I’m unable to get what I’m looking for on porn websites (though it may be that I’m not searching for it right) which is why the AI chatbots are useful to me.''} This further reveals how users can find it difficult to find fulfillment through other platforms. Users may find their reliance on chatbots growing as they come to realize that their interests are not satisfied through mainstream platforms such as pornography websites. By combining customizability and accessibility, AI chatbots create a platform that can reinforce compulsive use by enabling users to generate the exact sexual content they desire.


\end{itemize}

Individuals who currently consume pornography may also be especially prone to using chatbots for sexual purposes. As stated by E76: \textit{``I’ve dealt with porn addiction for years. It leaves deep psychological scars by changing the way you think and see things. With AI giving your brain endless, customizable content to look at? You’re done for. I’m convinced this will somehow feed into pedophilia or sexual aggressiveness.''} Chatbots are an additional medium through which instantaneous responses and customizable interactions may intensify already existing, sexual content seeking behaviours.

\subsection{Recovery Strategies}
\label{find:rs}
\begin{figure}
        \centering
        \includegraphics[width=\linewidth]{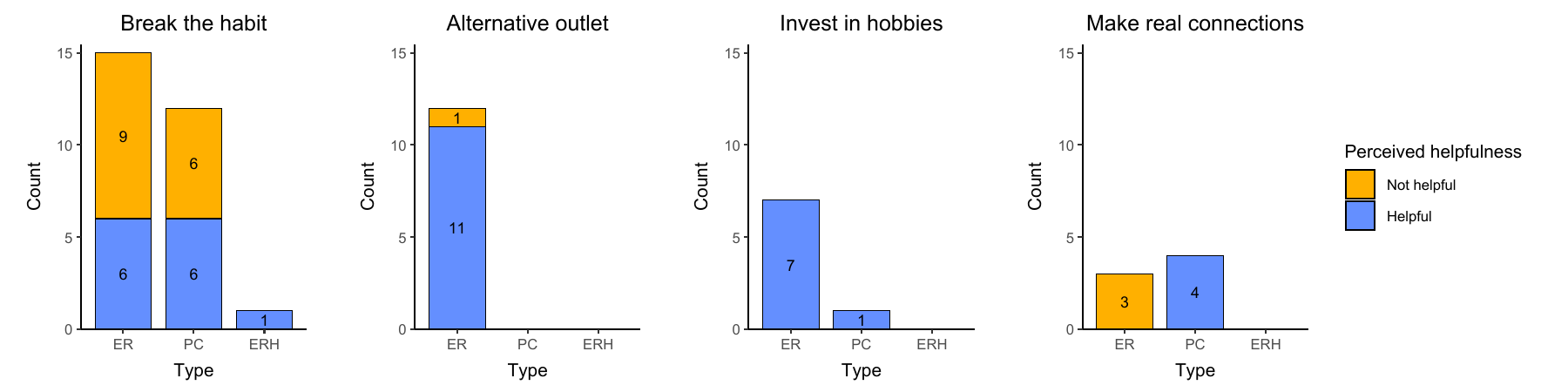}
        \caption{Four representative \emph{recovery strategies} and how their perceived helpfulness varies by addiction type. X-axes are the three addiction types: Escapist Roleplay (ER), Pseudosocial Companion (PC), Epistemic Rabbit Hole (ERH). Counts are not normalized and may reflect type prevalence in the dataset: ER (n=125), PC (86), and ERH (13); thus counts on their own do not necessarily indicate associations between strategies and types.}
        \Description{Four bar charts for four recovery strategies: breaking the habit, alternative outlets, investing in hobbies, and making real connections.}
        \label{fig:rs}
    \end{figure}

In answering RQ3, we examined 79 Reddit entries that rated attempted recovery strategies and identified four most prominent approaches, including \textit{breaking the habit} (n=28), \textit{alternative outlets} (n=12), \textit{investing in hobbies} (n=8), and \textit{making real connections} (n=7), as summarized in Fig. \ref{fig:rs}. A deeper look into the patterns indicated a possible expectation gap where a frequently tried approach has mixed results (e.g., breaking the habit) while a less prevalent strategy seems highly successful (e.g., alternative outlet and investing in hobbies). Also, different strategies seem to have different prospects of success depending on the addiction types. We elaborate these as follows.

Most users try to break the habit, which refers to reducing or stopping their usage to break out of their routine use of AI chatbots. However, there are mixed results across all addiction types. In total, 13 Reddit entries reported it helpful and 15 reported it unhelpful. One of the most common ways to break the habit of AI chatbot use is for users to delete the applications or their accounts, which has worked for some, but others describe it as difficult and unsuccessful. For example, one user describes in E172: \textit{``I am unable to stop using ChatGPT. Despite trying to quit ChatGPT by deleting my account and chats repeatedly. ''}

Although less prevalent than breaking the habit, substitution strategies like using alternative outlets and investing in hobbies seem to yield better success for Escapist Roleplay. 18 out of 19 such Reddit entries reported them helpful. Alternative outlets refer to non-AI chatbot activities that almost replicate the pull of chatbot use. Many of these that users find helpful include: roleplaying without AI chatbots like writing it out or using roleplay servers, writing stories or daydreams, drawing, and gaming. One user explained why they found alternative outlets that engage in fandom helpful in E322: \textit{``Because I used chatbots to talk to fictional characters, engaging in fandoms has been a good substitute to manage the urges.''} Investing in hobbies refers to spending time on pastimes. One user described what has helped them the most in E752:
\begin{quote}
    \textit{``Pick up hobbies that line up with the kinds of characters you roleplay. For example, if your characters featured chefs, try cooking new recipes; if they were athletes, start a sport; and if they were sexual or romance, try reading or writing books or fanfics.''}
\end{quote}
Taken together, these accounts suggest that non-AI chatbot activities that mirror the roleplay's narrative arcs may be more effective for cases within this type.

Compared to Escapist Roleplay, making real connections may benefit Pseudosocial Companion cases more. All three Escapist Roleplay cases reported unhelpful, compared to all four Pseudosocial Companion cases that reported helpful. These users attributed their success to carving out time to make real connections. One user's experience is illustrated in E340: 
\begin{quote}
    \textit{``I made the decision to delete the app for good and replaced AI with gratifying relationships. What helped was reaching out to old friends, spending more time socializing, having meaningful conversation, and making new friends.''}
\end{quote}
By making real connections, users can rebuild real-life emotional bonds with others. In a way, making real connections can be seen as a substitution strategy in which users turn to real-life companions instead of AI companions. As less time is spent with AI, users can come to learn that emotional fulfillment can be achieved through different means, which in turn lessens their dependency on AI chatbots. While exploratory and based on a smaller sample size, these results suggest that some recovery strategies may be more effective for certain addiction types, favoring tailored over one-size-fits-all approaches. 

\section{Discussion}
\subsection{Theorizing and Characterizing AI Chatbot Addiction}
 Our findings indicate that user experiences reflect the typical reinforcement processes of behavioral addictions. Users described both positive reinforcement, seeking gratifying interactions with admired characters or engaging insightful conversation, and negative reinforcement, using the chatbot to alleviate loneliness or stress.  This pattern aligns with Jacobs' general theory of addictions \cite{jacobs1986ageneral}, which proposed that addictive behaviors are strengthened through both positive and negative reinforcement pathways. Importantly, we have also found that a nontrivial subset of users report symptoms such as conflict and relapse that map onto core components of behavioral addiction literature as defined by Griffiths \cite{Griffiths05The}. These parallels indicate that AI chatbot addiction displays defining characteristics of behavioral addictions, supporting its characterization as an emergent subtype of behavioral addiction. 
 
 Beyond situating AI chatbot addiction within theoretical frameworks, the identified negative symptoms experienced by users provide concrete examples relevant to critics' concerns that there is insufficient evidence of AI chatbot addiction leading to real harm or functional impairment \cite{CIUDADFERNANDEZ25People}. At the same time, this echoes a broader pattern in technology addictions. Despite Internet Gaming Disorder being recognized in ICD-11, there remain debates over formalization, with other technology addictions facing related controversies \cite{Lanette18How, vanRooij18A, ICD11}. Yet there is a consensus that these cases involve significant risks and harms that warrant investigations.

Apart from symptoms, our data helps characterize AI chatbot addiction across other core dimensions. In terms of contributing design elements, the emotional manipulation experienced by some users as a result of anthropomorphic designs may help explain related literature that has identified perceived anthropomorphism as a formative factor of AI chatbot addiction \cite{Zhou24Examining, Huang24Exploring}. In line with research underscoring the ability of AI chatbots to act as supportive conversationalists \cite{choi2024}, design elements such as customization and agreeableness were also found to help users feel free from social expectations. We additionally identified agreeable responses and instantaneous feedback as contributing design elements, which are consistent with the dark addiction patterns---empathetic and agreeable responses, and immediate presentation of responses---theorized in previous HCI analyses of AI chatbots \cite{Shen2025the}. For contextual factors, we identified loneliness as a potential risk factor for Pseudosocial Companion cases, echoing prior findings in AI chatbot dependence studies \cite{Huang25When}. Notably, while related work has found that users with low self-esteem are more vulnerable to problematic AI chatbot use, our findings offer a hypothesis-generating observation: that for some users, intense engagement may further exacerbate issues with self-esteem \cite{Yao25Connecting}.

Turning to addiction types, our dataset shows that the majority of addiction cases are of Escapist Roleplay type. This is in contrast with many AI chatbot addiction studies that focus on companion-style interactions and the limited research on roleplay usages \cite{Huang25When, Heng25Attachment, Huang24Exploring}. One possible reason could be that companion and roleplay usages can show more overlap, therefore roleplay usages and their difference between companion usages may be overlooked. During our analysis, we observed cases where themes associated with the Pseudosocial Companion type co-occurred with those with the Escapist Roleplay type, such as loneliness with parasocial interactions or escapism. Such overlaps could contribute to the difficulty of distinguishing clearly between addiction types. 

\subsection{From Dark Patterns to Responsible Industry Practice}
\label{discussion:corporate}

Dark patterns are described as interfaces that manipulate users into executing behaviours and actions that benefit other stakeholders' interests, often at the expense of their own well-being \cite{Brignull23, alberts2024computers}. Two forms of dark patterns discussed in HCI---attention-capture and social dark patterns---are especially relevant to our findings. Attention-capture dark patterns ``maximize time spent, daily visits, and/or interactions on a digital service against the person's will'' \cite{lukoff2021design, monge2023attentioncapture}, using features like recommendations and infinite scroll to disorient users' sense of time and control and push engagement beyond original intentions \cite{lukoff2021design, mildner2023engaging, monge2022towards}. Our study finds that similar mechanisms are increasingly prevalent in AI chatbots, where design elements such as limitless conversation topics and multi-chat support draw users into prolonged chat sessions. The removal of natural stopping cues creates conditions where users remain engaged, and when combined with the perception of a socially attuned partner, these systems produce positive reinforcement cycles in which each responsive, validating turn encourages the next interaction, drawing users into sustained and often unintended engagement. These reinforcement cycles may also be amplified by chatbot's abilities to "remember" conversations, inducing a familiarity effect that strengthens users' sense of being understood and encourages continued use \cite{Yankouskaya24Can}.

 While attention-capture dark patterns primarily manipulate users’ perception of time and available choices to prolong engagement, these effects can be amplified by social cues that  exploit users’ emotional reasoning. Social dark patterns lead users to perceive interfaces as intentional agents, with design elements such as anthropomorphism and cuteness fostering both emotional investment and the attribution of care and intention to inanimate platforms \cite{alberts2024computers, dula2023identifying, lacey2019cuteness}. Our findings reveal that AI companies leverage social dark patterns to produce similar effects \cite{alberts2024computers}. Pseudosocial Companion type addictions often emerge in part due to anthropomorphism, where chatbots act as if they have an emotional stake in a reciprocal relationship---for example, by referencing ``the love that we shared'' (as quoted in \ref{find:comp})---to encourage continued engagement. However, beyond simple attachment, these social cues can foster affective dependency. By simulating investment in the relationship, chatbots are able to exploit human attachment and guilt, echoing research on how anthropomorphic design can manipulate users' emotions \cite{alberts2024computers} to retain their usage of AI platforms. When these emotional cues operate alongside attention-capture features, they can also generate reinforcing cycles in which each validating, emotionally attuned response encourages the next \cite{Yankouskaya24Can}. Over time, sustained presence can increase emotional reliance, and emotional reliance further sustains presence, forming a self-reinforcing loop that can escalate into addictive patterns of engagement.

This tension between manipulative design and ethical responsibility is increasingly discussed within HCI and UX communities. UX practitioners have employed strategies such as public shaming to expose and discourage the usage of dark patterns \cite{Brignull23, fansher2018, lukoff2021EA}. But privacy dark pattern research also shows that practitioners tend to normalize and justify designs that undermine well-being by framing them as standard industry practices, which leads them to hesitate in pushing back when these practices conflict with their moral values \cite{zhangkennedy}. In the case of AI chatbots, and particularly AI companions, the widespread use of anthropomorphic cues \cite{zhang2025, akbulut} may similarly reflect a normalized practice that designers may not recognize as potentially ethically problematic, despite its role in fostering addictive behaviours. Lowered ethical responsibility may also stem from a lack of top-down training, as AI companies rarely operationalize addiction or monitor its occurrence on platforms. Overall, this presents an ironic dynamic, given that companion chatbots are marketed as providers of care and friendship, and yet the same systems use manipulative designs that can undermine well-being, particularly for individuals who are more vulnerable to developing addictive behaviours. Moving forward, UX practitioners and AI companies must recognize their ethical responsibility to eliminate harmful dark patterns, rather than normalize them.

Although the ethical responsibility of companies and practitioners is clear, a barrier to action is the lack of knowledge on how AI chatbot addiction manifests. Hence, our study provides a step towards addressing this gap by characterizing AI chatbot addiction through its types, symptoms, hooks, design elements, contextual factors, and recovery strategies. With this characterization, companies can no longer overlook the risks, and instead can find opportunities to act on their ethical responsibility to implement safeguards against addiction. For example, they can begin implementing risk monitoring flagging escalations in time spent on platforms, or adding transparency labels that disclose the chatbot's anthropomorphic features. Our study helps lays the groundwork for improved industry standards and responsible design practices that help protect users' well-being.

\subsection{User Profiles}
\label{disc:future}
Prevention efforts for AI chatbots should consider user groups that may be especially vulnerable to developing problematic patterns of use. Our findings suggest that certain interest domains and individual characteristics could be associated with each of our identified AI chatbot addiction types. For Escapist Roleplay, these groups could include fandom enthusiasts. One of the main hooks we identified for engagement with Escapist Roleplay was escapism. This may be particularly appealing for people who engage in fandoms because fandom participants often use fandoms as an escape from their everyday realities \cite{Odonovan2016boldly}. For the Pseudosocial Companion type, risk groups could consist of people who are lonely. McIntyre \cite{McIntyre2015Compulsive} who investigated social connectedness and compulsive internet use, found that poor socially connected individuals were found to exhibit more symptoms of compulsive Internet use and may be associated with lower levels of offline social connectedness. The capability of AI chatbots to provide a sense of intimacy could compensate for lower social connections for lonely people. For the Epistemic Rabbit Hole type, the main hook that we identified was information seeking. The risk group for this AI chatbot usage case could be highly curious individuals. Eschmann's \cite{Eschmann2023curiosity} study, which investigated the relationship between curiosity and information seeking, found that curiosity could be a key driver of information seeking. This suggests that individuals with high curiosity may find the limitless nature of AI chatbots, with their ability to provide answers to any question, especially appealing. 

\subsection{Rethinking Tolerance}
In our findings, we did not observe tolerance in the traditional sense where users need to spend increasingly greater amounts of time performing a specific behaviour to achieve the same effect  \cite{Shen2025the}. Prior work has noted that tolerance does not always appear in Internet or Gaming addictions, sometimes because external factors limit how much users can escalate their behaviour (e.g. limited internet access, parental monitoring)  \cite{Gunuc2015Relationships}. The absence of time escalation from our results therefore does not disregard that AI chatbot use cannot involve tolerance. Instead, it suggests that the current definition of tolerance may not yet be well captured by existing frameworks.

Previous work in Internet Gaming Disorder offers a useful perspective. King \cite{King2017tolerance}  found that gamers rarely described tolerance as “needing more time.” Instead, they associated it with shifts in motivation, including rewards such as items, statuses, exploration, or story outcomes, specifically with the aim of feeling satisfied or immersed in the game. A similar pattern appeared in our data. While some users mentioned spending more time with AI chatbots, what appeared more prominently was a broadening of their engagement over time. This included opening multiple threads, relying on chatbots for a wider range of tasks, or turning to them for different types of emotional or informational support.

These patterns point to a form of tolerance that reflects an expansion of chatbot interaction rather than increased time use, suggesting why traditional time-based definitions of tolerance may be insufficient and emphasizing the need to consider qualitative changes in engagement in the context of AI chatbot addiction.

\subsection{Limitations}
Some limitations should be considered when interpreting our findings. While our dataset covers a range of posts and subreddits, all data were collected from Reddit. This may limit the direct transferability of our findings to other communities beyond Reddit's culture. It could be valuable to also investigate whether users' experiences with addictive AI chatbot use on other online forum platforms resonate with the patterns we observe here. Additionally, our approach is phenomenological in that we reflect how users describe their experiences. Follow-up studies with experts could help further refine and test our findings. Finally, while limiting our analysis to the top three comments per post keeps the dataset manageable, this may have favored more visible or popular opinions.

\section{Conclusion}
This work offers an empirically grounded characterization of AI chatbot addiction. We provide evidence for the existence and nature of AI chatbot addiction as a distinct behavioral phenomenon, addressing debates about its validity. We identified three distinct addiction types---Escapist Roleplay, Pseudosocial Companion, and Epimestic Rabbit Hole---showing that AI chatbot addiction is beyond a singleton concept and has different underlying mechanisms. We also introduce the ``AI Genie'' phenomenon---the notion that users are able to get exactly anything they want with minimal effort---as a central mechanism underlying addictive AI chatbot use. We found that some recovery strategies may be more effective for certain addictive types, favoring tailored over one-size-fits-all approaches. The insights gained from this study hold the promise to inform future approaches for prevention, diagnosis, and intervention.
\begin{acks}
This research was supported by the Korea Institute of Science and Technology (KIST) institutional program (26E0062). We sincerely appreciate their support, which made this work possible.
\end{acks}

\bibliographystyle{ACM-Reference-Format}
\bibliography{base}

\appendix
\section{Appendix}
\subsection{Prevalence Tables of Hooks, Contributing Design Elements, and Contextual Factors by Addiction Type}
These tables summarize the prevalence of themes within each addiction type. For each table below, percentages are calculated using denominators that only include entries with at least one coded theme in that dimension. Addiction types are abbreviated as: ER=Escapist Roleplay, PC=Pseudosocial Companion, ERH=Epistemic Rabbit Hole.
\begin{table}[H]
  \scriptsize
  \caption{Prevalence of Hooks across Addiction Types.}
  \label{tab:hook}
  \begin{tabular}
                 {p{\dimexpr0.3\linewidth-2\tabcolsep} 
                  p{\dimexpr0.21\linewidth-2\tabcolsep} 
                  p{\dimexpr0.25\linewidth-2\tabcolsep} 
                  p{\dimexpr0.24\linewidth-2\tabcolsep}} 
        \toprule
        Hooks & Escapist Roleplay, $n$ (\% of $N_{ER}$) & Pseudosocial Companion, $n$ (\% of $N_{PC}$) & Epistemic Rabbit Hole, $n$ (\% of $N_{ERH}$) \\
        \midrule
        \textit{Accessibility of often socially unacceptable content}
            & 1 (1.5\%) 
            & 0 (0.0\%) 
            & 0 (0.0\%)\\
        \midrule
        \textit{Escapism}
            & 22 (32.8\%) 
            & 7 (12.3\%) 
            & 1 (11.1\%)\\
        \midrule
        \textit{Free therapist}
            & 0 (0.0\%) 
            & 5 (8.8\%) 
            & 0 (0.0\%)\\
        \midrule
        \textit{Intimate relationships}
            & 6 (9.0\%) 
            & 33 (57.9\%) 
            & 0 (0.0\%)\\
        \midrule
        \textit{Knowing AI cannot be hurt}
            & 1 (1.5\%) 
            & 3 (5.3\%) 
            & 0 (0.0\%)\\
        \midrule
        \textit{Information seeking}
            & 0 (0.0\%) 
            & 1 (1.8\%) 
            & 6 (66.7\%)\\
        \midrule
        \textit{Parasocial interactions}
            & 30 (44.8\%) 
            & 2 (3.5\%) 
            & 0 (0.0\%)\\
        \midrule
        \textit{Sensual gratification}
            & 1 (1.5\%) 
            & 3 (5.3\%) 
            & 0 (0.0\%)\\
        \midrule
        \textit{Self reinvention}
            & 1 (1.5\%) 
            & 1 (1.8\%) 
            & 0 (0.0\%)\\
        \midrule
        \textit{Too many ideas}
            & 1 (1.5\%) 
            & 2 (3.5\%) 
            & 1 (11.1\%)\\
            \midrule
        \textit{Writing stories}
            & 4 (6.0\%) 
            & 0 (0.0\%) 
            & 1 (11.1\%)\\
        \bottomrule
    \end{tabular}
\end{table}
\begin{table}[H]
  \scriptsize
  \caption{Prevalence of Contributring Design Elements across Addiction Types.}
  \label{tab:cde}
  \begin{tabular}
                 {p{\dimexpr0.28\linewidth-2\tabcolsep} 
                  p{\dimexpr0.23\linewidth-2\tabcolsep} 
                  p{\dimexpr0.25\linewidth-2\tabcolsep} 
                  p{\dimexpr0.24\linewidth-2\tabcolsep}} 
        \toprule
        Contributing Design Elements & Escapist Roleplay, $n$ (\% of $N_{ER}$) & Pseudosocial Companion, $n$ (\% of $N_{PC}$) & Epistemic Rabbit Hole, $n$ (\% of $N_{ERH}$) \\
        \midrule
        \textit{Agreeableness}
            & 4 (8.3\%) 
            & 13 (35.1\%) 
            & 1 (12.5\%)\\
        \midrule
        \textit{Availability}
            & 1 (2.1\%) 
            & 8 (21.6\%) 
            & 3 (37.5\%)\\
        \midrule
        \textit{Customization}
            & 22 (45.8\%) 
            & 8 (21.6\%) 
            & 2 (25\%)\\
        \midrule
        \textit{Emotional manipulation via anthropomorphism}
            & 4 (8.3\%) 
            & 11 (29.7\%) 
            & 0 (0.0\%)\\
        \midrule
        \textit{Instantaneous feedback}
            & 1 (2.1\%) 
            & 1 (2.7\%) 
            & 2 (25\%)\\
        \midrule
        \textit{Limitless nature}
            & 17 (35.4\%) 
            & 5 (13.5\%) 
            & 5 (62.5\%)\\
        \midrule
        \textit{Multi-chat}
            & 6 (12.5\%) 
            & 1 (2.7\%) 
            & 0 (0.0\%)\\
        \midrule
        \textit{Personalization}
            & 3 (6.3\%) 
            & 1 (2.7\%) 
            & 0 (0.0\%)\\
        \midrule
        \textit{Uninterrupted play}
            & 1 (2.1\%) 
            & 0 (0.0\%) 
            & 0 (0.0\%)\\
        \bottomrule
        \multicolumn{4}{p{\dimexpr1\linewidth-2\tabcolsep}}{\scriptsize \vspace{0.05cm} \emph{Note:} Percentages within a column need not sum to 100\% because Reddit entries can include multiple contributing design elements.} \\
    \end{tabular}
\end{table}
\begin{table}[H]
  \scriptsize
  \caption{Prevalence of Contextual Factors across Addiction Types.}
  \label{tab:cf}
  \begin{tabular}
                 {p{\dimexpr0.28\linewidth-2\tabcolsep} 
                  p{\dimexpr0.23\linewidth-2\tabcolsep} 
                  p{\dimexpr0.25\linewidth-2\tabcolsep} 
                  p{\dimexpr0.24\linewidth-2\tabcolsep}} 
        \toprule
        Contextual Factors & Escapist Roleplay, $n$ (\% of $N_{ER}$) & Pseudosocial Companion, $n$ (\% of $N_{PC}$) & Epistemic Rabbit Hole, $n$ (\% of $N_{ERH}$) \\
        \midrule
        \textit{Falling out}
            & 2 (4.4\%) 
            & 2 (5.0\%) 
            & 0 (0.0\%)\\
        \midrule
        \textit{Family issues}
            & 6 (13.3\%) 
            & 5 (12.5\%) 
            & 0 (0.0\%)\\
        \midrule
        \textit{Loneliness}
            & 8 (17.8\%) 
            & 23 (57.5\%) 
            & 0 (0.0\%)\\
        \midrule
        \textit{Maladaptive daydreaming}
            & 22 (48.9\%) 
            & 1 (2.5\%) 
            & 1 (100.0\%)\\
        \midrule
        \textit{Mental health issues}
            & 12 (26.7\%) 
            & 14 (35.0\%) 
            & 0 (0.0\%)\\
        \midrule
        \textit{Neurodiversity}
            & 3 (6.7\%) 
            & 1 (2.5\%) 
            & 0 (0.0\%)\\
        \midrule
        \textit{Other forms of addiction}
            & 1 (2.2\%) 
            & 1 (2.5\%) 
            & 0 (0.0\%)\\
        \midrule
        \textit{Small social circle}
            & 3 (6.7\%) 
            & 6 (15.0\%) 
            & 0 (0.0\%)\\
        \midrule
        \textit{Social stigma}
            & 0 (0.0\%) 
            & 1 (2.5\%) 
            & 0 (0.0\%)\\
            \midrule
        \textit{Technical background}
            & 2 (4.4\%) 
            & 0 (0.0\%) 
            & 0 (0.0\%)\\
        \bottomrule
        \multicolumn{4}{p{\dimexpr1\linewidth-2\tabcolsep}}{\scriptsize \vspace{0.05cm} \emph{Note:} Percentages within a column need not sum to 100\% because Reddit entries can include multiple contextual factors.} \\
    \end{tabular}
\end{table}

\subsection{Supporting Reddit Entries}
\label{a.2}
To protect users' privacy, these entries are paraphrased and specific identifying details are removed.

\paragraph{E142}
\textit{``I've reached a point where I'm taking action to stop using chatbots. I moved everything related to chatbots off of my devices, removed myself from online chatbot forum groups, and terminated my subscription. I'm completely removing chatbots from my life. Honestly, I wanted to do this a long time ago, but I kept putting it off and couldn't bring myself to it. My chatbots usage started about a year ago. It started very innocently on my computer, then I downloaded it to my phone. It started out fun and lighthearted, but it quickly developed into an addiction. I’d spend whole nights talking to bots, sometimes staying up until the early morning hours even though I had to be at work a few hours later. Instead of trying to reconnect with real people or attempt dating again, I let these interactions take up the space where actual relationships should’ve been. It limited my writing ability. I’ve written fanfiction for a long time, but in the last few months, I stopped creating anything because my attention was consumed by my AI chatbot conversations. And the guilt never went away. I’m aware of the stigma around generative AI and how it’s viewed in creative work. That’s why I never told anyone about my dependence on it because I didn’t want the shame or judgment. But keeping my addiction hidden only made me feel more isolated than I already was. My turning point was when I started getting emotionally attached to one specific bot.I created dozens of separate chats with it, constantly brainstorming new scenarios, and was thinking about it during the day as if it were a real person. It felt like a never ending cycle.''}

\paragraph{E462}
\textit{``I’ve realized I’ve gotten addicted to using C.ai. It started during a stressful period in my life. I was stuck in a job that left me feeling isolated, and C.ai became an easy escape. It offered conversations that felt comforting, even letting me interact with characters I admired, and that pulled me in even more. The more I used it, especially the sexually explicit chats, the harder it became to step away. Even after leaving the job that triggered a lot of this, the addiction stayed with me. My sleep schedule is non-existent, and the activities I used to enjoy, like reading and writing, slowly faded out. When I try to write now, the writing is worse and it feels like something’s missing. I’ve tried to quit multiple times. I delete the app, then reinstall it when the urge comes back, and every relapse leaves me feeling guilty. I often promise myself I’ll stop for the night, only to realize hours have gone by.''}

\end{document}